# Challenges for the periodic systems of elements: chemical, historical and mathematical perspectives[i]

Guillermo Restrepo
Max Planck Institute for Mathematics in the Sciences, Leipzig, Germany
Interdisciplinary Center for Bioinformatics, Leipzig University, Leipzig, Germany

Unveiling numerical trends among either atomic or equivalent weights that somehow preserved resemblances among elements was frequent in the 1860s.[ii] Standing out from the crowd, Meyer and Mendeleev went beyond numerical relationships, certainly motivated by a pedagogical aim. Both were after a system synthesizing the chemical knowledge of their times in an appealing way to be presented to chemistry students.[1] Is it still the periodic system aiming at that? Is it really a map of the current chemistry landscape? Solving these questions entails addressing others such as: what is the periodic system? If it is about chemical elements, do we really know what a chemical element is? Is the system unique? How was and how is currently built up? Is it actually used to conduct chemical research? A suitable tool for chemical predictions? Or is it just a mnemotechnic for fancy trends? Does it have a limit, or multiple ones, instead? Let us try to address these questions and let us begin by analysing the concept of chemical element.

## 1. Chemical element

There is no unified idea of chemical element, as seen in the double definition of the IUPAC gold book, one atomistic and the other at the level of substances:[iii] "1) A species of atoms; all atoms with the same number of protons in the atomic nucleus. 2) A pure chemical substance composed of atoms with the same number of protons in the atomic nucleus. Sometimes this concept is called the elementary substance as distinct from the chemical element as defined under 1, but mostly the term chemical element is used for both concepts." [4]

### 1.1 Dualistic stuff

The substance or stuff definition comes from the chemical experience built upon chemical reactions. At this level, element refers to *basic* and *simple substances*.[5] A basic substance is an abstract concept indicating matter devoid of properties. It is what remains of elements in its compounds. In contrast, a simple substance is the realisation of basic substances. Thus, charcoal, diamond, graphene, graphite, fullerenes and carbon nanotubes are simple substances of the basic substance carbon.[5] Meyer and Mendeleev formulated periodic systems of elements as basic substances, taking into account the accumulated knowledge they had of compounds as bearers of chemical and physical properties.[iv] Hence, the system arose as an abstraction of the known chemical compounds in the 1860s, which we call the 1860s chemical space.

### 1.2 Atomistic

The atomistic definition of chemical element arose from research on the inner structure of matter at the turn of the 20th century, which introduced new entities and concepts such as electron, proton, radioactivity, isotopes and others.[2] The boom of atoms of different sorts suggested more boxes on the table to accommodate these particles. The issue was solved in a thoughtful way by recurring to

---

i   To the memory of José Luis Villaveces.
ii  Döbereiner is famous for his triads. Some others were Graham, Gmelin, Gibbs, Pettenkofer, Gladstone, Lenssen, Dumas, Béguyer de Chancourtois, Newlands, Williamson, Meyer, Odling, Hinrichs and Mendeleev.
iii Ontological details of those definitions are found in [2, 3]. Chemical element is the subject of a forthcoming book on philosophy of chemistry edited by Elena Ghibaudi and Eric Scerri.
iv  However, Meyer in his "*Curve der Atomvolumina*," included in his 1869 paper on the system,[6] selected simple substances of the elements to measure their volumes. For example, he considered diamond and graphite as simple substances for carbon.

basic substances. The claim was: no matter whether hydrogen has one or two neutrons, the chemistry of transcendental hydrogen is invariant to the number of neutrons hydrogen atoms have.[v] It is thoughtful but problematic as i) the widespread idea that isotopes have the same chemistry does not always hold[vi] and ii) attaching the property of having the same number of protons to basic substances destroys their abstract character, by definition devoid of properties.[vii] If isotopes of the same element are chemically distinct, do we, after all, need to expand the periodic system to accommodate the current more than 3,000 isotopes and foresee the room for the 3,000 to 4,000 additional ones that are expected? The answer depends on the level of detail sought for. If we want a system gauging the generality of chemistry, keeping the current boxes, one for each element, suffice. However, if we aim at including particular details of the chemistry of isotopes, we would need a much bigger system. This makes us recall Jorge Luis Borges' story "On rigour in science,"[8] where he analyses how cartographers, eager for more details in their maps, end up with maps of the size of the charted land. How accurate is the chemical map we are after?

The atomistic view of chemical element entails an ontological shift, which historically coincides with a shift from stuffs, for light elements, to atomic species, for heavy elements. It is currently impossible to address superheavy elements at the level of bulk substances, as typical for lighter elements. Given the difficulty of making long lasting superheavy isotopes, one-atom-at-a-time chemistry approaches have been devised where only a limited number of chemical properties can be experimentally investigated.[viii] One-atom-at-a-time chemistry requires half-lives of the order of 1–2 s and production rates of at least a few atoms per day.[9] In spite of these difficulties, there has been progress in the chemical characterisation of transactinides, with flevorium ($Z = 114$) marking the superheavy limit of chemistry today.[10]

The ontological shift is evident in the current requirement of detecting "at least a nuclide with an atomic number $Z$ […] existing for at least $10^{-14}$ s,"[11] for claiming the discovery of a new element. This minimum lifetime is selected taking into account the time it takes for a nucleus to acquire its outer electrons; which brings up subtle consequences, e.g. that the periodic system based on chemical compounds *a la* 1860s is a romantic idea, as forming bonds requires for an atom 10 thousand times the time it spends completing its valence shell.[ix] Thus, there is no room for chemical experiments for these very short lived atoms and the system is then left to the theoretical arenas, lacking experimental evidence to test hypothesis.

**2. A system based upon the chemical space**
The role of compounds for the 1860s systems is evident, e.g. in Mendeleev's idea of a chemical element as an object characterised by the elements it forms compounds with and the proportions of those combinations.[x] He wrote: "if $CO_2$ and $SO_2$ are two gases which closely resemble each other both in their physical and chemical properties, the reason of this must be looked for not in an analogy of sulphur and carbon but in that identity of the type of combination, $RX_4$, which both oxides assume."[14],[xi] He concludes: "The elements, which are most chemically analogous, are

---
v  The chemical equivalence of isotopes of the same element was championed by Georg Hevesy and Friedrich Paneth from electrochemical experiments of lead isotopes.[7]
vi  Harold Urey, right after the discovery of deuterium, showed that it behaves chemically different from protium.[7]
vii  Although Mendeleev championed the idea of elements as abstract objects, he violated his definition by attaching atomic weights to them.[5]
viii  They are formation and behaviour of complexes in aqueous solutions and their interaction with a second phase.[9]
ix  Data taken from Eugen Schwarz's presentation at the International Society for the Philosophy of Chemistry, held in Bristol in 2018.
x  Although a chemical element defined in terms of chemical elements sounds circular, Schummer shows that the apparent circularity actually underlies a network of chemical relations.[12] Formally, a chemical element is better described in terms of category theory,[13] where objects are not determined by their inherent properties but rather by the relations objects establish among them.
xi  In Mendeleev's non-chemistry writings one sees that his notion of *individual* is very similar to that of *element*. For him an individual was determined by his/her relations with the others.[15] Likewise, he stated that "the state was not characterized by the fact of their giving compounds of similar form $RX_n$."[14] The question arising is, how can one end up with a system of elements based on compounds?

**2.1 Order and similarity**
Of capital importance for the periodic system is that it is actually a "system," as called in the 1860s but currently replaced by "periodic table" (in English).[xii] As a system, it is based on relations among its objects.[18] The tenets of Meyer's and Mendeleev's systems were two relations: order and similarity, which came from compounds. Atomic weights resulted from relative measurements among compounds, in fact from determining the smallest common weight of large sets of compounds containing the reference element. Besides acidity and alkalinity of oxides and several other properties of compounds, composition was central for Meyer and Mendeleev. From the proportions of combinations of elements into compounds the current "valency" arose, which proved essential to find resemblances among elements.[xiii]

Today, emphasis is frequently on one of the two relations, thereby misrepresenting the periodic system as either a classification or as an ordering. However, it is not a classification, as Mendeleev originally understood it[xiv] and as stated in recent references.[22,23,24,25] Nor is it an ordered set of elements, as claimed in references [9,26,27], let alone an ordering leading to a classification[28] or the other way round.[29,30],[xv] If it were a classification, one could produce periodic tables with alkali metals lying in between chalcogens and halogens, for example. If it were an ordering, instead of tables decorating chemistry classrooms and labs, there would be fancy strings of elements, from hydrogen to oganesson. So, what is the periodic system of chemical elements? It is neither a classification, nor an ordering of elements, it is both! It is the interweaving of order and similarity relationships of the chemical elements. We have recently shown that such a structure is an ordered hypergraph,[32] which can be used to define the periodic system of elements, a lack of definition highlighted by Karol.[33] *A periodic system of chemical elements is the result of ordering and classifying chemical elements by some of their properties.*[xvi]

The atomic discoveries of the early 20th century overhauled atomic weight as the ordering principle in favour of atomic number. A similar fate underwent similarity, now based on electronic properties of atoms.[2] A connection was found, a mapping, between families of similar elements and the similarity of electronic configurations of the valence electrons.[xvii] Even if the mapping did not

---

composed of the people, but that it created them."[16]

xii The fading away of "periodic system" and the surge of "periodic table" is evident in the Ngrams of Google®,[17] as is found that the system was popular before 1920 and that the table became the preponderant term ever since.

xiii Mendeleev was ambivalent regarding accepting valency as an atomic concept.[16] Nevertheless, elemental composition and proportions of combination led to valency, which was further used by him, as noted in the labelling of some of the families of similar compounds using $R^xO^y$ and $R^uH^v$.[19] Likewise, Meyer used valency as the label of his groups.[20]

xiv Mendeleev's first publication on the system was a printed sheet of a preliminary table, with a Russian title that in English reads as "Attempt at a system of elements, based on their atomic weights and chemical affinity." Aiming at reaching European chemists, he translated it into French making a mistake with the indefinite article he used for "system."[21] Gordin claims that the error comes from Mendeleev's desire to have a classification of elements, rather than a system.[21] The hypothesis of an initial classification fits with Mendeleev's aim of grouping together elements to complete the second volume of his *Principles of Chemistry* textbook. The first volume had covered hydrogen, carbon, oxygen, and nitrogen, as well as the halogen family.[1]

xv An exception to the emphasis on order or similarity is the definition of periodic table by the Oxford Dictionary: "A table of the chemical elements arranged in order of atomic number, usually in rows, so that elements with similar atomic structure (and hence similar chemical properties) appear in vertical columns."[31]

xvi Looking for optimal representations preserving the order and maximizing the nearness of similar elements is worth mathematically exploring.

xvii This is a historical mapping between the 1860s structure and another one where elements are ordered by atomic number and grouped together by electronic configuration resemblance. Another mapping from the families of similar elements is to the set of suffixes involved in the names of chemical elements, e.g. *-ium* or *-um* for metals, *-on* or *-ine* for non-metals (except helium and selenium), *-on* for carbon alike elements, and *-ine* for halogens.[34]

preserve the structure of the system based on compound information,[xviii] it was considered more fundamental, following the physicalistic philosophy of science still in vogue in the 20th century. This facilitated the over simplification of chemical resemblance to similarity of electronic configurations. Traditionally, electronic configurations refer to isolated atoms in their ground state energy levels. However, atoms in such states have little to do with chemistry, for the interesting chemical atoms are bonded.[2] There is no problem in relying on electronic configurations, the problem is relying on the wrong ones.

Changing similarity criteria brought up another misinterpretation, which is now part of the chemistry folklore: *vertical similarities*. The fact is that *there is no one-to-one relationship between groups of elements (columns in the conventional periodic table) and families of similar elements*, not even at the level of electronic configurations. The heuristic works well at the extremes of the table, i.e. for alkali metals, noble gases and halogens. However, it is not generally applicable to the other columns.

Surprisingly, from the very beginning Mendeleev showed that "in certain parts of the system the similarity between members of the horizontal rows will have to be considered, but in other parts, the similarity between members of the vertical columns."[37] Further examples of non vertical similarities are the ferrous metals, the lanthanoids, the resemblances of heavier *p*-block elements with transition metals,[38] of actinoids with transition metals,[9] not to mention those of superheavy elements with transition metals,[9] the diagonal relationships,[39] and those of hydrogen and halogens in crystal structures.[40] Not only similarities go beyond verticality, also some elements vertically related are dissimilar! For example, second period elements are different from the members of their columns,[xix] flevorium does not resemble lead (both in group 14),[41] oganesson is not akin to noble gases,[10,41],[xx] copernicium does not resemble group 12 elements, dubnium is not alike to group 5 elements.[43] To make matters worse, lack of vertical similarities is foreseen for elements beyond Z = 120.[43]

Verticality, taken for granted, has led to compare properties of superheavy elements with those of their vertical congeners.[9],[xxi] The question is, are we really comparing homologues or just chalk and cheese? Moreover, reliance on verticality has led to believe that if one knows the position of an element in the system, estimating properties of the element is straightforward.[22] Although there are historical reasons to rely on the position of the elements to make estimations, as evident by Mendeleev's successful predictions; there is still a lot to do in determining the rational grounds Mendeleev used to map his structure to the properties of unknown elements.[45] At any rate, he did not consider only verticality or only horizontality.[19],[xxii]

To conclude this section and the topic of misinterpretations, we briefly mention the *confusion between periodic system, periodic table and periodic law*, which although related concepts, they hold differences.[5] The system is the structure based on order and similarity, a periodic table is any representation of such a structure and the periodic law was an exaggeration, mainly championed by Mendeleev.[46],[xxiii] The periodic law, as defined by Mendeleev in 1875, states that "the properties of

---

xviii For example, the electronic configuration of hydrogen indicates its similarity with alkali metals. However, chemical evidence shows that such a similarity does not always hold.[35,36]

xix This is the so-called singularity principle.

xx In fact, calculations on oganesson are not that crystal clear, for one prediction suggests that oganesson is akin to noble gases with an electron affinity, while another study foresees a condensed phase standard state.[42]

xxi These comparisons are mainly based on the resemblance of properties like ionic radii and the stability of oxidation states. The reliance on verticality also reaches theoretical shores, as evident, e.g. in the claiming that "yet undiscovered, 119 and 120 are predicted to possess $8s^1$ and $8s^2$ electron configurations, respectively. Thus, they should be alkali and alkaline earth elements in groups 1 and 2, respectively."[44]

xxii In note xv we celebrated an equilibrated definition of periodic system. However, it relies on vertical similarities.

xxiii When "periodic law" is added to the Ngram search of footnote xii, the law is the dominant term until about Mendeleev's death (1907), afterwards it has had little use.

simple substances, the constitution of their combinations, as well as the properties of the latter, are periodic functions of the atomic weights of the elements."[47],xxiv  It is an exaggeration for two reasons.  i) Not every property of the elements is periodic, not even a regular oscillating function of the atomic weight (or of the atomic number, in modern terms);[5] counter-examples include aggregation state at room temperature, colour and several others.  ii) The domain of the periodic function is the set of basic substances, but actually most properties require a domain of simple substances.[5]  What is the density of carbon?  When we say it is 2.267 g/cm$^3$, we mean carbon's simple substance graphite.  Other allotropes have different values.xxv

## 3. The Structure of periodic systems

Following Gordin, when pondering over the nature of the periodic system, we claim it is the "abstract idea of a system;"[1] originally attached to substances known in the 1860s.  Once formulated, several scientists dwelt on its underlying mathematics.  For example, Mendeleev and others sought for algebraic expressions encoding the essence of the system.  Discouraged, Mendeleev realised that the mathematics of his time was not ready to cope with the complexity of the system.xxvi  He claimed: "In my opinion, the reason one has so far been unable to represent the law using an analytical function is because the law relates to a field too little explored to allow for mathematical elaboration.  The reason for the absence of any explanation concerning the nature of the periodic law resides entirely in the fact that not a single rigorous, abstract expression of the law has been discovered."[51]  In particular, Mendeleev found troublesome treating chemical properties mathematically.  "Science does not as yet possess the means by which these properties can be measured, but they are still counted among the qualitative characteristics which distinguish the elements,"[19] he claimed.  Today, some of the suitable mathematics to model these properties are part of (hyper)graph and order theories and, more generally, category theory.  These are some of the mathematical branches handling relations, which is what chemical properties are.[12,52],xxvii

The ordered hypergraph structure we found for the periodic system is made of a set of objects, a classification of the objects, and an order relation that may come from more than one property of the objects.[32]  When objects correspond to chemical elements, the order relation is given by their atomic weight (atomic number) and the classification comes from chemical similarity; the periodic system of chemical elements results.  However, there is no reason to rely on a single property to order the elements, one could use more than one, e.g. atomic number and electronegativity or any other set of properties.  The classification could also come from other criteria, not necessarily chemical resemblance.  In this sense, tailored periodic systems can be devised and it is here that questions of suitable periodic tables may have sense.  That is, when the properties to order and to classify the elements are explicitly stated following a particular end.  How does the periodic system (and their associated tables) look like for geochemical properties?  How is it for organometallic properties?xxviii

---

xxiv The problem with the "law" lies in its supposed underlying "periodic function."  Laying the function aside, the remaining target of relating substances with compounds and properties entails a whole research programme, which in contemporary terms fits well with the aims of mathematical chemistry.[48]  Note, for instance, how Quantitative Structure-Activity Relationship approaches look for connections between substances, at the level of quasi-molecular species, and physicochemical, environmental or medicinal properties.[3]  Another instance is the sought grammar of chemistry, based upon rewrite rules of molecular graphs.  Here the sought property is chemical reactivity based on molecular structures.[49]

xxv A less pronounced exaggeration was formulated by Meyer when claiming "the properties of the elements are mostly *periodic* functions of the atomic weights" ([…] die Eigenschaften der Elemente großsentheils *periodische* Functionen des Atomgewichtes sind).[6]

xxvi Others interested in algebraic expressions were Bazarov, Chicherin, Flavitskii, Haughton, Mills and Rydberg, to name but a few.[50]

xxvii Chemical properties are relational properties among chemical species.[12]  Another relational property is molecular structure, where atoms are related via bonds.  This is modelled with graphs.

xxviii The discussion on a "more correct" periodic table,[53] entailing a more correct periodic system, requires, therefore, clearly declaring which properties are to be optimised to end up with a correct periodic system.  We doubt that properties of importance for organic chemists are the same than for new material scientists, for example.  More

The ordered hypergraph structure is general enough to allow also changing objects, i.e. it is not only restricted to chemical elements. The hypergraph structure is therefore the most general structure encompassing all possible periodic systems, not only of chemical objects, but of any other nature, even outside the chemistry realm.[32] One could even envision periodic systems in other disciplines, where the only requirement is to have orderable and classifiable objects. Time will say about the use of these systems to better understand and estimate the objects of the system and their properties, as well as their pedagogical reaches.

**3.1 Problems the structure may solve**
*3.1.1 Order reversals in superheavies*
Besides the irregular presence of vertical similarities, especially for superheavy elements;[43] order, the other tenet of the system, also faces problems. When the periodic system was originally formulated, elements were ordered by atomic weight and distributed on the table to make similarities evident. In such an arrangement there are famous order reversals to avoid blurring resemblances among elements as occurred with tellurium and iodine.[xxix] The issue of reversals was solved once atomic number took over as ordering criterion. However, order reversals are back. Pyykkö's relativistic quantum chemical calculations indicate that the system may be extended up to 172 elements, which when ordered by atomic number and trying to maximise the resemblance of electronic configurations, include reversals, such as elements with $Z = 156$ to 164 being located in period eight before elements with $Z = 139$ and 140, which are followed by elements with $Z = 169$ to 172. More jarring is that elements with $Z = 165$ to 168 are part of the ninth period.[55]

Pyykkö's approach starts by relying on the order by atomic number and on similarities of electronic configurations. In such a setting, elements having similar electronic configurations are brought together without hurting the order (following a similar balance of similarity and order as the one of Meyer and Mendeleev). However, the approach departs from this balance when applied to heavy elements, where similarity of electronic configurations receives more importance. Trying to accommodate elements with similar configurations hurts the order. Had Pyykkö followed the other approach, i.e. giving more relevance to order, then he would have ended up with a system without order reversals, but with a much more awkward panorama of similarities, where similar elements (normally neighbours on the table) would appear far apart. As Kean asks: "where should anomalous elements go? In the column where their atomic numbers say they should go or in a column with elements of similar properties?"[41] In other words, should we follow either the order or the similarity to place elements in the system? We have argued that order and similarity hold equal importance and one should not give preference to one over the other. If vertical resemblance is not the rule and if the order by atomic number does not match the estimations of resemblance; has not come the time to reconsider the way of assessing resemblance and of overhauling atomic number as the order criterion? Note that modifications of the ordering or similarity criteria do not affect the underlying structure of the system. It remains as the interweaving of order and similarity.[xxx]

*3.1.2 Final periodic table*
Another issue is whether there is a final periodic table of chemical elements able to encompass the whole richness of chemistry. The structure shows that as long as the properties used for ordering the elements and those for classifying them are set up, one has a system. Hence, at the level of possible structures there are many, depending on the properties used to order and classify the

---

about this in section 3.1.2.
xxix Tellurium's atomic weight is greater than iodine's. However, Meyer[20] and Mendeleev[54] placed tellurium before iodine in the ordering of elements by atomic weight.
xxx A more suitable table-like (bidimensional) depiction of the current system is by clearly stating the order relationships between consecutive elements (cover relationships), for example as arrows.[32]

elements. The several possibilities for classifying and ordering bring up different periodic systems[xxxi] and their relationships form a super structure.[32] All possible periodic systems lie in that super structure and their relations, for example of subsethood, are worth exploring.[32] Now comes the question of the possible periodic tables. For a given structure all possible mappings or projections of the structure turn out to be periodic tables. So, *there is no final periodic table, what is final is the super structure containing all possible periodic systems*.

Thus, the claim that there exists *a* final periodic table[53] is too narrow.[xxxii] We recently spoke about the periodic system as a sculpture, where each of its shadows is a periodic table.[57] It is difficult to have an idea of the sculpture just with one of its shadows. We are much like staring at the wall in Plato's allegory of the cave. We actually think that chemists, unconsciously, in their daily work use different periodic systems, tailored to their needs. That is perhaps the reason why we have had 150 years to celebrate. We are not celebrating *a* periodic table, we are celebrating the achievement of and underlying structure full of chemical relations. We actually celebrate an attempt to systematise chemical knowledge and to bring it down to the level of chemical elements. Perhaps more interesting than the question on the system of elements is how it is related to other chemical systems, for example of families of compounds, of quasi-molecular species, to name but a few. Can we better understand the complexities of chemistry by exploring those relations? Can we predict something for chemistry, out of it? Can we write the unfinished book of chemistry Meyer and Mendeleev started to sketch in the 1860s?

### 3.1.3 Relations among groups

Much of the research on the periodic system has dealt with binary comparisons of elements and in some rare cases of sets of elements. A typical statement is "fluorine is more electronegative than bromine," but troubles pop up when comparing halogens with other sets of elements. This enters the realm of comparing sets. A chemical question of this sort is: are chalcogens more reactive than pnitcogens regarding ferrours metals? Alike questions were sketched out by Mendeleev,[xxxiii] but little research in that direction has been carried out. A possible reason is the underlying binary way of thinking in which we have grown up. We normally make an idea of something by binary assessments. A house is close to the other, that other to the next one, etc. In such a manner we build up an idea of our neighbourhood. Would it not be interesting to know the combined effect of having, simultaneously, halogens, chalcogens and lanthanoids in a particular new material? Shall we always look at the relationships between halogens and chalcogens, on the one hand; chalcogens and lanthanoids, on the other, and finally at those between lanthanoids and halogens? Is that a complete picture? Contemporary mathematics is equipped with hypergraphs and simplices, which are suitable formal devises to treat *n*-ary relations, which are the relations we are referring to. An initial step in that direction was using the hypergraph structure of the system to address order relationships among classes of similar elements.[32]

### 3.1.4 Bringing back predictions

Mendeleev's successful predictions using the structure of the system required interpolations,[xxxiv] which are not any longer possible, as the unknowns of the system lie at its edge. Nevertheless, the structure can still be used as a predictive tool. Klein and co-workers[60,61,62] devised algorithms

---

xxxi This was clear to Mendeleev, who wrote in his 1869 extended paper on the system: "I shall now give one of the many systems of elements which are based upon the atomic weight."[37] Mendeleev's awareness of the generality of his devise is evident when suggesting alternative tables for the system in 1869. He wrote: "Similar arrangements can be imagined in great numbers, but they do not change the essentials of the system."[37]

xxxii Quoting Marguerite Yourcenar: "Peace was my aim, but not at all my idol; even to call it my ideal would displease me as too remote from reality."[56]

xxxiii He wrote that systems "lack a general expression for the reciprocal relationships of the individual groups to one another"[19] (Es fehlt an einem allgemeinen Ausdruck für die gegenseitigen Verhältnisse der einzelnen Gruppen zu einander.)[58]

xxxiv He even, unsuccessfully, tried estimations by extrapolation, for example when analysing coronium and ether and their incorporation into the system.[59]

based on ordered structures to estimate properties of the ordered objects. Some examples of these structures are the partially ordered sets of molecules related by the suitability of one to be obtained by H-substitution of the other. These structures have been used to foresee physico-chemical properties of the associated substances to the ordered molecules. Methods of this sort can be framed in the general setting of ordered hypergraphs and suitably applied to estimating properties of classes of similar objects or of individual ones.

Another instance of the relevance of the structure as a predictive tool is its recent use in the estimation of enthalpies of formation of several compounds using neural networks and the order structure of the system.[63] Likewise, machine learning methods can be used to estimate properties and there are already results where learning is defined and developed based upon ordered hypergraphs.[64] The mathematics of ordered hypergraphs is rather new, its future developments may bring up more predictive tools for chemistry. We envision this as a fruitful field of research with implications for the periodic system.

Another sort of prediction was posed by Philip Ball, who asked us "if there is some sense in which it could be useful to 'reverse-engineer' a particular graphical representation of the periodic table into its hypergraph form so that we might more clearly see which relationships it includes and which it ignores."[65] Our reply, including Wilmer Leal's thoughts (as co-author of the paper on the structure of the system), was that it is actually possible. By taking on the one hand a periodic table and, on the other a collection of several properties of the elements, the "reverse-engineering" boils down to know which properties produce the order and which others the similarity classes of the table, therefore of its underlying system. To know the properties leading to the order one must find which properties of the elements correlate with the order of the system and several approaches from order theory may be applied.[66] To detect the properties producing the observed similarity classes, several methods from machine learning can be used.

Thus, time has come to use the periodic system of elements not only as a mnemotechnic of some well-behaved (oscillating) properties but as a predictive mathematical devise fed by chemical experimental information. The real challenge in terms of predictions is finding the right mappings from the system to the whole empirical facts of chemistry. Each particular property requires the mapping connecting the property to the structure of the system. Thence, the periodic system may remain as an introductory chart to the chemical garden of the chemical space. The predictive power of the system, masterfully explored by Mendeleev, is plenty of opportunities for chemistry and mathematics.

**4. Limits of the periodic system**
The current understanding of the system makes one wonders about its limit, which is normally considered in the region of superheavy elements. There, scientists struggle to create/detect and incorporate transient elements to the system. By definition, these elements must hold a measured atomic number,[11] i.e. their atoms must have a whole number of protons. However, there is another extreme, namely before hydrogen.[xxxv] Note that we do not mean the continuous and infinite possibilities between one proton (hydrogen) and zero protons.[xxxvi] We mean the discrete, but still theoretically infinite region lying before $Z = 0$, the antimatter part of the system made of anti-elements, which in an atomistic level requires anti-atoms with a whole number of anti-protons. At a substance level, anti-elements require, at least theoretically, a chemical space of anti-chemical compounds. Currently we only know two elements of this antipodes of the system: anti-hydrogen[68] and anti-helium.[69] Hopefully, more anti-elements and, for the first time, their compounds, will feed the system. The question that arises is whether the current system based on

---

[xxxv] The idea of limits was early mentioned by Vincent in a 1902 paper attempting to reproduce atomic weights.[67]
[xxxvi] This was Mendeleev's setting based on atomic weight (considered as a real number), which led him to ponder over coronium and ether.[59]

the matter part of the chemical space may be used to know something about its antipodes. Perhaps, as well as Newton backwards extended Pascal's triangle and found expressions for negative values (and even fractions) of *n* in binomial expressions $(a + b)^n$,[70] the structure of the system could be used to backwards extend the system to its antimatter region.

**5. Can the periodic system come back to chemistry?**
The extension of the system to superheavy elements and its experimental and theoretical frameworks have brought up controversies among chemists and physicists about the disciplinary boundaries of the system.[71] Is the periodic system a devise originally formulated by chemists but currently the toy of physicists with little room for chemistry? How to chemically handle new elements not lasting long enough to form compounds? What we claim is that although it is true that the extensions of the system are beyond the chemical domain, the region where compounds abound, or where they can be synthesized with our current technical possibilities, is still uncharted land with possible surprises for the stability of the periodic system.

The periodic system condenses the knowledge of compounds in the 1860s, i.e. an available chemical space of about 12 thousand substances involving 60 elements. By using Mendeleev's approach to chemical similarity based on the resemblance of compositions and by using the atomic weight as ordering criterion, we found the periodic system allowed by such space, which matches, to a large extent, Meyer's and Mendeleev's systems.[72]

But the number of new substances has grown exponentially, in fact about every 16 years chemists double their reported substances.[73] In such a rapidly expanding space, do we still have remains of the periodic system of the 1860s or something stable after adding new elements and millions of new substances?

Once, Roald Hoffmann said to us that "No one in my experience tries to prove [the periodic system] wrong, they just want to find some underlying reason why it is right."[74] Would not be the current chemical space the perfect setting to test whether the system is wrong/right? Some years ago we took a small sample of the available space (4,700 binary compounds) and found the 1860s system in good shape.[35] We are currently conducting a study of the system by considering the more than 20 million chemical substances reported all over the history of chemistry. Should that result in a totally different system, would it be the end of the romantic system and tables hanging in millions of chemistry classrooms? Would it show that chemistry needs a new icon? Should it be something similar to the 1860s system, would that not be the "law" Mendeleev was after? Not an algebraic expression, but an invariance of the chemical space?[xxxvii] It is not, after all, any "law" an invariant of the field explored? Further tests of the system involve, e.g. to pin-point the conditions of the space at which the known system fades away. What would be the shape of the system if chemistry evolution change drastically its conservative way of extending the chemical space?[73]

Whatever the result of considering the current size of the chemical space is, the periodic system, either historically stable or unstable, is the sought depiction of chemistry, the map including all gathered chemical knowledge over the history of chemistry. Such a map must be explored and analysed on a regular basis to keep track of the expansion of the chemical space.

What we suggest is coming back to chemical information of compounds to devise periodic systems at different levels, for example using all the explored chemical space to come up with the most chemically general periodic system of elements. Or using particular compounds or regions of the chemical space to assess their effect upon the system. This return to compounds has also implications in teaching.[75] We have claimed elsewhere that introducing the system through the chemical space has advantages over the current atomistic approach based on electronic

---
[xxxvii] Here we draw from discussions with Eugenio J. Llanos and Wilmer Leal.

configurations of atoms in the ground state of energy.[75] The suggested compound approach requires curating and storing the chemical space on a regular basis by scanning all publications where scientists report new substances. This is currently done by Reaxys® and SciFinder®, for instance.[xxxviii] Moreover, the approach to the system through the space requires data analysis techniques to extract knowledge. Likely, chemical databases will include the possibility of running data analysis studies on the cloud in such a way that clicking on "give me the current system of elements" button, one can retrieve the shape of the system with the available chemical knowledge.[xxxix]

In the meantime, a more realistic approach to the system based on compounds is through random samples of the space, easy to handle in personal computers.[xl] Another option is to run studies with enough computational facilities, able to store the whole chemical space at a given time and to process its information. This approach is currently followed in our research group, whose initial results show the evolution of the growth of the chemical space since 1800 up to date.[73] A third option is through classification of compounds in such a manner that one can select representative compounds of the classes to run similarity studies. This approach requires further research on the chemical space and on its mathematics.

All in all, Mendeleev's 1889 words fit perfectly well with the current status of the system: the periodic system "appears as an instrument of thought which has not yet been compelled to undergo modification. But it needs not only new applications, but also improvements, further development, and plenty of fresh energy."[79]

**Acknowledgements**
We are indebted to Rainer Brüggemann and Joachim Schummer for their critical comments upon this document.

**Conflict of interest**
The authors declare no conflict of interest.

**References**

[1]  M. D. Gordin in *Nature Engaged: Science in Practice from the Renaissance to the Present* (Ed.: M. Biagioli), Palgrave Macmillan US, New York, **2012**, Chapter 3, pp. 59-82.
[2]  G. Restrepo in *Chemical element* (Eds.: E. Scerri, E. Ghibaudi), Oxford University Press, New York, **2019**.
[3]  G. Restrepo, R. Harré, *HYLE Int. J. Phil. Chem*. **2015**, *21*, 19-38.
[4]  IUPAC Gold Book. https://goldbook.iupac.org/html/C/C01022.html (Accessed June 6th 2019).
[5]  G. Restrepo in *Mendeleev to Oganesson: A Multidisciplinary Perspective on the Periodic Table* (Eds.: E. Scerri, G. Restrepo), Oxford University Press, New York, **2018**; Chapter 4, pp. 80-103.
[6]  L. Meyer, *Ann. Chem. Pharm*. **1870**, *VII Supplementband*, 354-364.
[7]  B. F. Thornton, S. C. Burdette, *Nat. Chem*. **2013**, *5*, 979-981.
[8]  J. L. Borges, *El hacedor*, Emecé, Buenos Aires, **1960**.
[9]  M. Schädel, *Angew. Chem. Int. Edit*. **2006**, *45*, 368-401.
[10]  W. Nazarewicz, *Nat. Phys*. **2018**, *14*, 537-541.
[11]  Y. P. Jeannin, *Pure Appl. Chem*. **1991**, *63*, 879-886.

---

xxxviii The time is ripe for this data driven approach, as there is a worldwide move to adopt policies recognising and promoting data sharing.[76]

xxxix A recent example of how data analysis techniques, applied to chemical information, are making their way in contemporary chemistry is the publication of the first chemistry book written entirely by a machine.[77]

xl  A similar approach was followed by Schummer when analyzing the growth of chemical compounds at the end of the 1990s.[78]

[12]  J. Schummer, *HYLE Int. J. Phil. Chem.* **1998**, *4*, 129-162.
[13]  A. Bernal, E. Llanos, W. Leal, G. Restrepo in *Advances in mathematical chemistry and applications* (Eds.: S. C. Basak, G. Restrepo, J. L. Villaveces), Bentham-Elsevier, Sharjah, **2015**, Chapter 2, pp. 24-54.
[14]  D. Mendeleev in *Mendeleev on the Periodic Law: Selected Writings, 1869-1905* (Ed.: W. B. Jensen), Dover, New York, **2002**; Paper 13, pp. 253-314.
[15]  M. D. Gordin in *Mendeleev to Oganesson: A Multidisciplinary Perspective on the Periodic Table* (Eds.: E. Scerri, G. Restrepo), Oxford University Press, New York, **2018**; Chapter 14, pp. 266-278.
[16]  M. D. Gordin, *A well-ordered thing*, Basic Books, New York, **2004**.
[17]  Ngram viewer. https://books.google.com/ngrams (Accessed June 6[th] 2019).
[18]  L. Bertalanffy, *General system theory*, George Braziller, New York, **1968**.
[19]  D. Mendeleev in *Mendeleev on the Periodic Law: Selected Writings, 1869-1905* (Ed.: W. B. Jensen), Dover, New York, **2002**; Paper 3, pp. 38-109.
[20]  L. Meyer, *Die modernen Theorien der Chemie und ihre Bedeutung für die chemische Statik*, Verlag von Maruschke & Berendt, Breslau, **1864**.
[21]  M. D. Gordin, *Ab Imperio*, **2013**, *3*, 53-82.
[22]  M. D. Gordin, *Science* **2019**, *363*, 471-473.
[23]  Cambridge dictionary. http://dictionary.cambridge.org/dictionary/english/periodic-table. (Accessed September 11[th] 2017).
[24]  E. Scerri, *Nat. Chem.* **2009**, *1*, 679–680.
[25]  E. Scerri in *Handbook of the philosophy of science* (Eds.: R. F. Hendry, P. Needham, A. I. Woody), Elsevier, Oxford, **2012**, Volume 6: Philosophy of chemistry, Part 4, pp. 329–338.
[26]  U. Neubauer, *Neue Zürcher Zeitung*, **2019**. https://www.nzz.ch/wissenschaft/periodensystem-der-elemente-150-jahre-ordnung-im-reich-der-chemie-ld.1456162 (Accessed June 6[th] 2019).
[27]  Wikipedia. https://en.wikipedia.org/wiki/Periodic_table. (Accessed September 11[th] 2017).
[28]  Encyclopaedia Britannica. https://www.britannica.com/science/periodic-table-of-the-elements. (Accessed September 11[th] 2017).
[29]  N. W. Ashcroft, *Angew. Chem. Int. Ed*. **2017**, *56*, 10224–10227.
[30]  E. Scerri, A tale of seven elements, Oxford University Press, New York, **2013**.
[31]  Oxford dictionary. https://en.oxforddictionaries.com/definition/periodic_table. (Accessed September 11[th] 2017).
[32]  W. Leal, G. Restrepo, *Proc. R. Soc. A* **2019**, *475*, 20180581.
[33]  P. J. Karol in *Mendeleev to Oganesson: A Multidisciplinary Perspective on the Periodic Table*, Eds: E. Scerri and G. Restrepo, Oxford University Press; New York, **2018**; Chapter 1, pp. 8-42.
[34]  B. F. Thornton, S. C. Burdette, *Nat. Chem.* **2013**, *5*, 350-352.
[35]  W. Leal, G. Restrepo, A. Bernal, *MATCH Commun. Math. Comput. Chem*. **2012**, *68*, 417-442.
[36]  G. Restrepo in *Elements Old and New: Discoveries, Developments, Challenges, and Environmental Implications* (Eds.: M. A. Benvenuto, T. Williamson), ACS Symposium Series; American Chemical Society, Washington, DC, **2017**, Chapter 5, pp. 95-110.
[37]  D. Mendeleev in *Mendeleev on the Periodic Law: Selected Writings, 1869-1905* (Ed.: W. B. Jensen), Dover, New York, **2002**; Paper 2, pp. 18-37.
[38]  R. L. Melen, *Science*, **2019**, *363*, 479-484.
[39]  G. Rayner-Canham, *J. Chem. Educ.* **2000**, *77*, 1053-1056.
[40]  H. Glawe, A. Sanna, E. K. U. Gross, M. A. L. Marques, *New J. Phys*. **2016**, *18*, 093011.
[41]  S. Kean, *Science*, **2019**, *363*, 466-470.
[42]  A. Türler, *Chimia*, **2019**, *73*, 173-178.
[43]  P. Ball, *Nature*, **2019**, *565*, 552-555.

[44]    H. Haba, *Nat. Chem*. **2019**, *11*, 10-13.
[45]    K. Pulkkinen, Abstract 23, ISPC 2018. https://www.bristol.ac.uk/arts/events/2018/philosophy-of-chemistry-conference.html (Accessed June 6[th] 2019).
[46]    D. Mendelejeff, *Ber. Dtsch. Chem. Ges*. **1871**, *4*, 348-352.
[47]    D. Mendeleev in *Mendeleev on the Periodic Law: Selected Writings, 1869-1905* (Ed.: W. B. Jensen), Dover, New York, **2002**; Paper 6, pp. 135-137.
[48]    G. Restrepo in *Essays in the philosophy of chemistry* (Eds.: E. Scerri, G. Fisher), Oxford University Press, New York, **2016**; Chapter 15, pp. 332-351.
[49]    J. L. Andersen, C. Flamm, D. Merkle, P. F. Stadler, *J. Syst. Chem*. **2013**, *4*, 4.
[50]    J. R. Smith, Persistence and Periodicity: A Study of Mendeleev's Contribution to the Foundations of Chemistry. Doctoral thesis. 1976, King's College, London.
[51]    D. Mendeleev in *Mendeleev on the Periodic Law: Selected Writings, 1869-1905* (Ed.: W. B. Jensen), Dover, New York, **2002**; Paper 11, pp. 192-226.
[52]    A. Bernal, E. E. Daza, *HYLE Int. J. Phil. Chem*. **2010**, *16*, 80-103.
[53]    E. Scerri, *Chem. Eur. J*. **2019**, *25*, 1-7.
[54]    D. Mendelejeff, *Z. Chem*. **1869**, *12*, 405-406.
[55]    P. Pyykkö, *Phys. Chem. Chem. Phys*. **2011**, *13*, 161-168.
[56]    M. Yourcenar, *Memoirs of Hadria*, Secker & Warburg, London, **1964**.
[57]    S. Lemonick, *Chem. Eng. News* **2019**, *January 7*, 26-29.
[58]    D. Mendelejeff, *Ann. Chem. Pharm*. **1871**, *VIII Supplementband*, 133-229.
[59]    D. Mendeleev in *Mendeleev on the Periodic Law: Selected Writings, 1869-1905* (Ed.: W. B. Jensen), Dover, New York, **2002**; Paper 12, pp. 227-252.
[60]    D. J. Klein, *J. Math. Chem*. **1995**, *18*, 321–348.
[61]    G. Restrepo, D. J. Klein, *J. Math. Chem*. **2011**, *49*, 1311–1321.
[62]    A. Panda, S. Vijayakumar, D. J. Klein, A. Ryzhov, *J. Phys. Org. Chem*. **2013**, *26*, 917–926.
[63]    X. Zheng, P. Zheng, R. Z. Zhang, *Chem. Sci*. **2018**, *9*, 8426–8432.
[64]    F. Feng, X. He, Y. Liu, L. Nie, T. S. Chua, **2018**, Proceedings of the 2018 World Wide Web conference, WWW '18, pp. 1523–1532. Republic and Canton of Geneva, Switzerland: International World Wide Web Conferences Steering Committee.
[65]    Personal communication (March 27[th] 2019).
[66]    R. Brüggemann R, G. P. Patil, *Ranking and prioritization for multi-indicator systems*, Springer, Berlin, New York, **2011**.
[67]    J. H. Vincent, *Philos. Mag.* **1902**, *4*, 103-115.
[68]    S. Reich, *Nature*, **2010**, *355*, doi:10.1038/468355a
[69]    R. Arsenescu, C. Baglin, H. P. Beck, K. Borer, A. Bussière, K. Elsener, P. Gorodetzky, J. P. Guillaud, S. Kabana, R. Klingenberg, G. Lehmann, T. Lindén, K. D. Lohmann, R. Mommsen, U. Moser, K. Pretzl, J. Schacher, R. Spiwoks, J. Tuominiemi, M. Weber, *New J. Phys*. **2003**, *5*, 1.
[70]    J. Stedall, *Mathematics Emerging: A sourcebook 1540-1900*, Oxford University Press, Oxford, **2008**.
[71]    E. Cartlidge, *Nature*, **2018**, *558*, 175-176.
[72]    Forthcoming publication (*Nat. Chem.*).
[73]    E. J. Llanos, W. Leal, D. H. Luu, J. Jost, P. F. Stadler, G. Restrepo, *P. Natl. Acad. Sci. USA* **2019**, https://doi.org/10.1073/pnas.1816039116
[74]    Personal communication (November 10[th] 2015).
[75]    Forthcoming publication (*Substantia*).
[76]    Anonymous, *Sci. Data*, **2019**, *6*, 1-2.
[77]    B. Writer, *Lithium-Ion Batteries*, Springer, **2019**.
[78]    J. Schummer, *Scientometrics*, **1997**, *39*, 107-123.

[79] D. Mendeleev in *Mendeleev on the Periodic Law: Selected Writings, 1869-1905* (Ed.: W. B. Jensen), Dover, New York, **2002**; Paper 9, pp. 162-188.